\documentclass[conference]{IEEEtran}
\IEEEoverridecommandlockouts
\usepackage{cite}
\usepackage{amsmath,amssymb,amsfonts}
\usepackage{algorithmic}
\usepackage{graphicx}
\usepackage{textcomp}
\usepackage{xcolor}
\usepackage{float}
\usepackage{hyperref}
\usepackage{subcaption}

\usepackage{graphicx}
\usepackage{adjustbox}

\usepackage{cancel}
\usepackage[normalem]{ulem}
\usepackage{amsmath}
\usepackage{xcolor,cancel}
\usepackage{comment}

\newcommand{\heading}[1]{\noindent\textbf{#1}}

\def\BibTeX{{\rm B\kern-.05em{\sc i\kern-.025em b}\kern-.08em
    T\kern-.1667em\lower.7ex\hbox{E}\kern-.125emX}}
\begin{document}

\title{Using explainable AI to investigate electrocardiogram changes during healthy aging - from expert features to raw signals}

\author{Gabriel Ott, Yannik Schaubelt, Juan Miguel Lopez Alcaraz, Wilhelm Haverkamp, Nils Strodthoff* \thanks{GO, YS, JMLA and NS are with Oldenburg University, Oldenburg, Germany (email: \{gabriel.ott,yannik.schaubelt,juan.lopez.alcaraz,nils.strodthoff\}@uol.de. WH is with Charité Universitätsmedizin Berlin, Berlin, Germany (email: wilhelm.haverkamp@dhzc-charite.de). GO and YS contributed equally. Corresponding author: Nils Strodthoff.}
}

\maketitle

\begin{abstract}

Cardiovascular diseases remain the leading global cause of mortality. Age is an important covariate whose effect is most easily investigated in a healthy cohort to properly distinguish the former from disease-related changes. Traditionally, most of such insights have been drawn from the analysis of electrocardiogram (ECG) feature changes in individuals as they age. However, these features, while informative, may potentially obscure underlying data relationships. In this paper we present the following contributions: (1) We employ a deep-learning model and a tree-based model to analyze ECG data from a robust dataset of healthy individuals across varying ages in both raw signals and ECG feature format. (2) We use explainable AI methods to identify the most discriminative ECG features across age groups.(3) Our analysis with tree-based classifiers reveals age-related declines in inferred breathing rates and identifies notably high SDANN values as indicative of elderly individuals, distinguishing them from younger adults. (4) Furthermore, the deep-learning model underscores the pivotal role of the P-wave in age predictions across all age groups, suggesting potential changes in the distribution of different P-wave types with age. These findings shed new light on age-related ECG changes, offering insights that transcend traditional feature-based approaches.

\end{abstract}

\section*{Introduction}

\heading{Characterizing healthy aging through ECG changes} Cardiovascular diseases continue to represent the leading cause of mortality worldwide \cite{doi:10.1016/j.jacc.2022.11.005}. Analyzing the effect of healthy aging on the cardiovascular system enables to distinguish between an old but healthy and a younger but cardiovascular-constrained heart.
This is especially difficult as there exists a discrepancy between biological age and cardiovascular age \cite{Pavanello2020}. Thus, knowing what changes in the heart during healthy aging can help to avoid deaths since it enables early treatment through premature detection of cardiovascular diseases. These changes are most commonly assessed through changes in electrocardiogram (ECG) features of healthy people with age.

\heading{Shortcomings of prior work} However, by working on the feature level, relationships in the ECG that are not covered by any features may be excluded from the analysis. Previous research suggests that deep-learning models can outperform feature-based classifiers on age prediction with ECG data \cite{Zvuloni2023}. Studies like \cite{Attia2019,strodthoff2020deep} also used deep-learning models to infer the age from ECG data but they did not restrict their dataset to healthy people only. Moreover, when deep-learning models were used to do age prediction on ECG data of healthy subjects by \cite{similar_paper}, it was not analyzed how the models managed to detect the age and thus they failed to provide information on what changes in the heart during aging. Furthermore, \cite{age_pred_paper}, \cite{age_pred_paper_focus_hrv} and \cite{age_paper_focus_children} focused on analyzing what changes in the heart with age, but did not make use of deep-learning models.

\heading{Research questions} 
In this work, we aim to use techniques from explainable AI (XAI), to identify both features from feature-based age classifiers as well as ECG segments from deep learning models operating on raw data that are most important to discriminate between different age groups within a collective of healthy people. To this end, we address the following research questions: (1) Do both approaches produce insights that correspond to literature results? (2) Can this approach be used to discover new ECG feature correlations in the context of healthy aging across diverse groups?

\heading{Main contributions and findings}
We analyze a dataset \cite{dataset_paper,Goldberger2020:physionet} of 1,120 ECG recordings of healthy people with varying ages using two different models: an XResNet50 and an eXtreme Gradient Boosting (XGBoost) model. The XResNet50 operates on raw ECG data, while the XGBoost model uses long-range and short-range ECG features as input. Both models were trained to predict the age of a healthy person from their 1-lead-ECG and achieve a competitive performance (macro-AUCs of 0.73 and 0.77, respectively). After training, both models were then investigated with explainable AI methods where the XGBoost-model was analyzed with SHAP to find the most important ECG features for classifying the age. On the other hand, for the XResNet50 heartbeat-based saliency maps were superimposed to show the areas of interest for the age prediction task.

To summarize, our contributions and findings in this study are:
(1) Our approach reaches competitive performance even though it leverages very different feature sets. (2) The XGBoost model mainly leverages long-range features. The inferred breathing rate from the ECG declines with age for healthy people and that very high SDANN5 (average standard deviation of normal-to-normal RR-intervals within a 5-minute interval) values are more likely to be from a person aged 50 or more than from a person aged 34 or less, even though SDANN values generally decline with age. (3) By construction, the XResNet is only able to leverage short-range features. Given the XResNet insights, we show that the XResNet50 exploits relationships in the P-wave with age, presumably indicating that the distribution of different types of P-wave changes with age.

\section{Materials and methods}
\subsection{Background}
\heading{ECG analysis} 
ECG analysis is of great importance in understanding the impact of aging on heart health and mortality. As highlighted by \cite{Brant2023}, accelerated heart-aging, as indicated by ECG-age, is associated with a significant increase in all-cause mortality, underscoring the crucial role of ECG as a biomarker for cardiovascular risk. Investigating how the ECG of a healthy heart changes with age is therefore essential for mortality prediction. Furthermore, \cite{age_paper_focus_children} examined ECG features to assess changes in the autonomic nervous system across various age groups, revealing clear age-related trends. While their study demonstrated such trends, it utilized a limited set of ECG features. For a comprehensive understanding of ECG analysis and its diverse applications, including critical steps like preprocessing, feature extraction, selection, transformation, and classification, \cite{Merdjanovska2022} provides an informative survey that encompasses the breadth of this field.

\heading{Age prediction from ECGs}
\cite{similar_paper} used deep learning techniques on the automatic aging dataset \cite{dataset_paper} for age prediction. However, their approach involved reducing the original 15 age classes to only 4, which improved model performance but limited the potential for explainability insights. Meanwhile, \cite{Attia2019} successfully predicted the age of individual subjects with a notable average error of 7 years. It is worth noting that their dataset included individuals with various health conditions, raising the possibility of age inference from age-related diseases. A similar result was achieved on a public dataset in \cite{strodthoff2020deep}. \cite{Zvuloni2023} introduced a promising predictive approach comparing deep learning models working on raw ECG data and tree-based classifiers using ECG features. These models were trained on a substantial dataset of over 2.3 million 12-lead ECGs for diverse tasks. However, the study did not explore the explainability of their models, which remains an important aspect for further investigation in age prediction from ECG data. Our research addresses these gaps by considering a finer granularity of age groups, allowing for more detailed insights from explainable methods. Additionally, by focusing solely on data from healthy individuals, we ensure cleaner data and more reliable insights and hence less confounding factors, thus contributing to a more comprehensive understanding of age prediction from ECGs.

\heading{ECG explainability}
A recent review \cite{kalyakulina2023explainable} highlights the potential of using techniques of explainable AI (XAI) to uncover mechanisms underlying age prediction models, an approach that resonates very well with the approach taken in this work.
In the domain of ECG analysis, the significance of interpretability has gained prominence in recent studies, as nicely reviewed in recent systematic reviews \cite{Ayano2022}. Researchers have been actively integrating XAI techniques to enhance the interpretation of ECG data. However, in many cases, this crucial component get reduced to anecdotal evidence obtained from the straightforward application of commonly used attribution methods to handpicked examples to underline the validity of the proposed algorithm. On the contrary, two recent dedicated works on interpretability in the ECG domain \cite{Bender2023,wagner2023explaining} highlight the methodology of aggregated attributions across patients or entire patient populations, a technique that is also supposed to be used in this study.

\subsection{Dataset and data preparation}

\heading{Dataset} The Autonomic Aging dataset \cite{dataset_source,Goldberger2020:physionet} aims to quantify changes in cardiovascular autonomic function during healthy aging. It contains ECG recordings of 1,120 healthy-control subjects sampled at 1,000 Hz in a resting state under controlled measurement conditions. Nevertheless, for the purpose of this study, we considered only ECGs where the age information is given, leaving 1,095 patients. The patients' ages range from 18 to 92 years, and the recordings span from 8 to 35 minutes, with a mean of 19 minutes. Two different devices were used to measure the ECGs, a 1-lead and a 2-lead ECG recorder. Therefore, we used only the matching lead (II) from both devices. The gender distribution within the dataset shows a slight imbalance towards female patients (675 female and 420 male patients). The dataset contains 38 ECG recordings with missing values, however, most of them with only a few time steps across the complete signal at 1000 Hz, therefore, missing values were removed without excluding recordings. For approaches based on raw time series data, we work at a temporally downsampled resolution of 100~Hz, which was found to be sufficient for common diagnostic tasks \cite{Mehari2023S4}, and also removed a negligible number of missing values in the time series. 

In the dataset, all subjects are diagnosed as healthy, so we sampled 3-second crops in order to capture at least one complete heartbeat. Due to the imbalanced nature of the dataset, we opted for a 60/20/20 split between training, validation, and test sets at the subject level. Importantly, we maintained the age-group distribution in each set as far as possible while simultaneously ensuring that every age group was represented in all three sets. For consistency, all models presented in this work use identical splits for their training, validation, and test data. Table \ref{tab:dataset_composition} contains additional descriptive statistics of the processed dataset. See our source code for the dataset pre-processing steps \cite{Ott_ECG-aging_2023} for more details.

\begin{table}[ht]
\centering
\caption{{\bf Summary of the dataset composition.}}
\resizebox{\columnwidth}{!}{
\begin{tabular}{|l|c|c|c|}
\hline
          & Train & Validation & Test \\ \hline
Subjects  &     &    &  \\ 
Male      & 251 (38.20\%) & 84 (38.35\%) & 85 (38.81\%) \\ 
Female    & 406 (61.80\%) & 135 (61.65\%) & 134 (61.18\%) \\ \hline
Age       &     &   &   \\ 
Median      & 25-29y  & 25-29y & 25-29y \\ 
Quantile 25\% & 20-24y & 20-24y & 20-24y \\ 
Quantile 75\% & 35-39y & 35-39y & 35-39y \\ \hline
Patient ECG record in seconds &     &   &   \\ 
Median  & 920.74  & 922.51 & 916.3 \\ 
Quantile 25\%  & 900.3 & 900.23 & 899.65 \\ 
Quantile 75\%  & 1207.7 & 1222.5 & 1204.89 \\ 
Mean  & 1077.39  & 1112.9 & 1053.24 \\ 
Std. Dev. & 349.72 & 387.24 & 343.15 \\ 
Min  & 599.7 & 480.5 & 600.2 \\ 
Max  & 2056.7  & 2051.66 & 2168.2 \\ \hline
Total crops & 235947 & 81241 & 76886 \\ \hline
\end{tabular}
}
\label{tab:dataset_composition}
\end{table}

\begin{figure}
    \centering
    \includegraphics[width=\columnwidth]{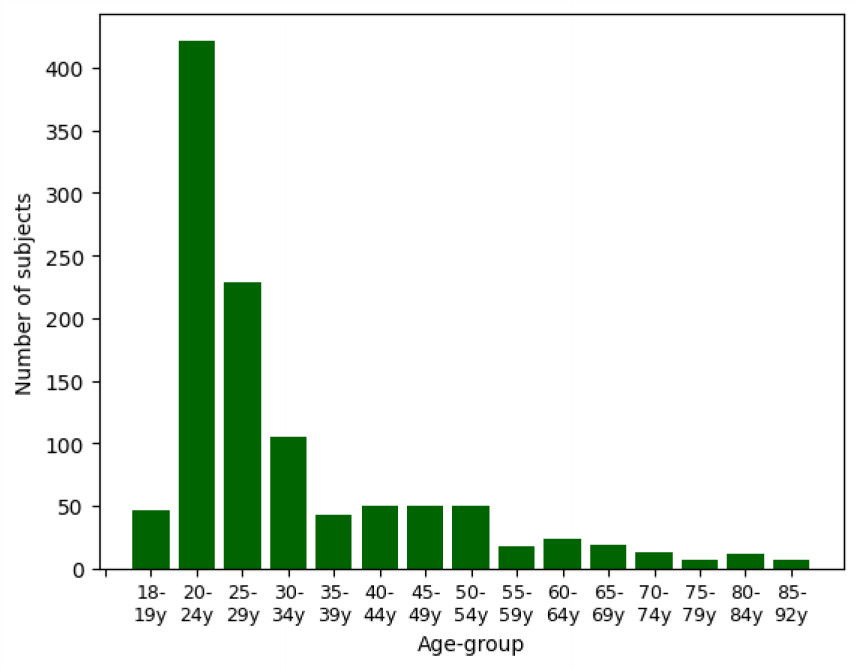}
    \caption{{\bf Age-group distribution.}
Age-group distribution in terms of age groups provided in the Autonomic Aging dataset\cite{dataset_paper}. The age groups span a range from 18 to 92 years, where the majority of patients are between 20 to 50 years old.}
    \label{fig:1}
\end{figure}

\heading{Age-group distribution} Fig \ref{fig:1} shows the age distribution across the dataset in terms of 15 age groups, where the first age group contains subjects aged 18 to 19, whereas all following age groups but the last cover age intervals of 5 years. There is a clear imbalance in the age distribution, with the majority in age group 20-24 with 422 samples, followed by age group 25-29 with 105 samples. On the contrary, the last four classes represent only 39 samples or correspondingly 3.4\% of the full dataset. Furthermore, it is important to mention that there are no male samples available for the age groups 75-79 years and 85-92 years. As past studies did not indicate a strong interaction effect between gender and age prediction, and dividing the dataset by gender would worsen the imbalance in the smaller age groups, we decided to ignore gender as a covariant in this study.


\subsection{Models and feature sets}

\heading{Overview} For the purpose of this study, we investigate two different classifiers working on two different feature sets but both predicting the subject's age: a residual neural network (XResNet50) operating on raw time series data and a tree-based model gradient boost decision tree classifier (XGBoost) operating on derived features. We carried out diverse experiments to find sufficient feature sets and preprocessing settings which could lead us to a scenario of better performance results. In each of the settings, we report test set scores of models selected based on held-out validation set scores. To facilitate continued research, we release the source code underlying our study \cite{Ott_ECG-aging_2023}.

\subsubsection{XGBoost} 

For the XGBoost model, we include long-range heart-rate variability (HRV) and short-range (SR) features. The HRV features describe how the ECG signal varies over time, and it contains features such as Standard Deviation of NN Intervals (SDNN), Root Mean Square of Successive Differences (RMSSD) and low and high-frequency powers to name a few. The HRV features were extracted from the ECGs with NeuroKit2 \cite{NeuroKit2_paper}. The SR features were calculated from fiducial points and comprise features such as R-R intervals, heart rate, peak amplitudes, and waves such as Q, R and S, and P and T respectively. The SR features were extracted per heartbeat with the python HeartPy toolkit \cite{heartpy_paper} in combination with NeuroKit2. To produce an age prediction for a whole ECG, the heartbeat-interval SR features were averaged over each ECG recording. This enabled us to combine the SR and HRV feature sets. As there are two different feature sets, all three combinations were tested: each alone and both sets combined. 

\heading{Training}
As a countermeasure against the label imbalance in the training dataset, we created artificially balanced datasets by oversampling the minority classes with a random oversampling technique. Lastly, regarding the model training, we performed a grid search to determine optimal hyperparameters based on validation set performance. After this process, the only hyperparameters where we found deviations from default values to be beneficial were max\_depth=10, max\_leaves=10, learning\_rate=0.008. For the explainability analysis, we leverage SHAP values \cite{lundberg2020local}. This is in line with a recent comparative study \cite{mehari2023ecg} where SHAP values showed good overlap with cardiologists' expert features.

\heading{Relevant features} At this point, it is worthwhile explicitly highlighting a number of ECG features that will play an important role for the later analysis:
\begin{itemize}
\item \textit{SDNN and SDNN5:} SDNN represents the standard deviation of normal-to-normal RR-intervals, while SDANN5 denotes the average SDNN calculated within a 5-minute interval. Similarly, SDANN1 refers to the average SDNN computed within a 1-minute interval. 

\item \textit{HRV\_PAS:} This metric quantifies the percentage of NN intervals within alternating segments, where NN intervals represent the time intervals between normal R-peaks in the ECG signal. 

\item\textit{Alpha-Features:} Alpha-features are derived from detrended fluctuation analysis (DFA) and provide insights into the auto-correlation between heartbeats. Specifically, alpha1 characterizes short-term correlations, while alpha2 captures long-term correlations in heart rate variability. 

\item\textit{pNN:} pNN20 signifies the percentage of heartbeat intervals with more than a 20-millisecond deviation from the previous interval, while pNN50 represents the corresponding percentage for intervals with more than a 50-millisecond deviation. 

\item\textit{MCVNN:} MCVNN stands for the median absolute deviation of RR intervals divided by the median of RR intervals, providing valuable information about heart rate variability.
\end{itemize}
These ECG features serve as critical components for our analysis, and understanding their definitions is essential for comprehending the subsequent sections of this paper.

\subsubsection{XResNet50}
The XResNet50 deep learning model works with raw time series data. We chose the XResNet50 model, which showed competitive performance with the best-performing convolutional neural networks for a range of different ECG classification tasks \cite{strodthoff2020deep,NilsNet}. It represents a one-dimensional adaptation of a commonly used ResNet-type convolutional neural network from computer vision \cite{he2019bag}. Here we additionally restrict to a single input channel as appropriate for 1-lead ECG data. 

\heading{Training} The XResNet50 was trained with the AdamW optimizer and weight decay \cite{loshchilov2018decoupled}. We investigate two different loss functions, namely focal loss (FL) \cite{lin2017focal} and cross-entropy loss (CEL). The learning rate was set to $10^-5$ for FL and $10^-2$ for CEL and adjusted with a reduced learning rate on the plateau scheduler, which divided the learning rate by 10 if the loss did not decrease for 2 consecutive epochs. We trained on 20 epochs with early stopping after 3 consecutive epochs. Oversampling the training set produced insufficient performance results in early experiments. Thus, as a countermeasure against the unbalanced training set, class weights are applied instead. The class weights were set to the inverse of each age group's number of occurrences in the training set. We investigate different scenarios, firstly by using two different loss functions and, secondly, by applying training class weights.

We trained the models on crops of 3s length and aggregate predictions from multiple crops using mean output predictions to obtain sample-level predictions, see \cite{Mehari2023S4}
for a detailed analysis of the benefits of this procedure. Note that using 3s-crops limits the XResNet50 to detect short-range patterns. We leverage the methodology proposed in \cite{ecg_xai_preprint} to compute beat-aligned attribution maps over entire patient subgroups. In particular, we use saliency maps as attribution maps as saliency was the only attribution method that satisfied the sanity checks proposed in \cite{ecg_xai_preprint}.

\subsection{Performance metric}
For comparability with earlier works, we report accuracy as a performance metric but stress the severe shortcomings of accuracy in the presence of severe class imbalance as is the case here. As the main performance metric, we report the macro-averaged (over age groups) area under the receiver operating curve (macro-AUC), which is less affected by class imbalance and operates on output probabilities rather than dichotomized outputs. To assess the uncertainty of our predictions due to the finite size of the test set, we resort to bootstrapping on the test set. We report  2.5 and 97.5 percentiles of the test set scores, i.e.\ 90\% confidence intervals for the test set scores. We indicate these within brackets behind the point estimate for the score such as $0.5 \text{\footnotesize{ (0.025, 0.975)}}$.

\section{Results}

\subsection{Predictive performance results}

\subsubsection{XGBoost}

In Table \ref{tab:xgboost_configuration_results}, we present the performance evaluation of the XGBoost model operating on different feature sets, including SR, HRV, and a combined feature set HRV+SR, trained on both balanced (oversampled) and unbalanced (original) datasets. The results indicate a nuanced performance pattern, with the balanced configuration generally exhibiting slightly superior performance compared to the unbalanced setting. Notably, the model operating on SR features performed worse with an AUC of 0.70. Conversely, the model incorporating both SR and HRV features and trained on a balanced dataset demonstrated the highest efficacy, achieving an AUC score of 0.77 on the test set. The performance gain over the model leveraging on HRV features confirms that the combined model actually exploits both short-range as well as long-range features.

\begin{table}[!htt]
\centering
\caption{
{\bf XGBoost model macro-AUC on the test set for different configurations.}}
\begin{tabular}{|r|l|l|}
\hline
 & Balanced & Unbalanced \\ \hline
HRV & 0.74 \scriptsize{(0.69, 0.79)} & 0.73 \scriptsize{(0.68, 0.77)}\\ \hline
SR & 0.70 \scriptsize{(0.63, 0.75)} & 0.70 \scriptsize{(0.65, 0.74)}\\ \hline
\textbf{HRV+SR} & \textbf{0.77 \scriptsize{(0.72, 0.80)}} & 0.72 \scriptsize{(0.68, 0.77)}\\ \hline 
\end{tabular}
\label{tab:xgboost_configuration_results}
\end{table}

To set these results into perspective, we also show a direct comparison between our feature-based XGBoost model and a previously introduced feature-based approach \cite{similar_paper}. For comparability, we follow their approach and consolidate the original 15 age groups into 4 broader categories. Both models demonstrate closely aligned accuracy scores, with our XGBoost model achieving 0.684 (95\% CI: 0.62-0.74) and the prior feature-based model at 0.688 (95\% CI: 0.64-0.73). The similarity in performance with almost identical point estimates and largely overlapping confidence intervals underscores the parity between our XGBoost model and the established feature-based approach. This reinforces the reliability of our findings and underscores the suitability of our model for age group classification tasks, laying the foundation for further explainability investigations.

\subsubsection{XResNet}

Table \ref{tab:xresnet-aucs} presents the performance evaluation for different XResNet50 configurations. Notably, the experimental findings underscore the superiority of focal loss over cross-entropy loss as the preferred choice for this task with a severely imbalanced label distribution. Furthermore, irrespective of the loss function employed, it is evident that the model attains significantly enhanced performance levels when trained on the unbalanced training dataset. The most noteworthy configuration emerges with focal loss applied to the unbalanced training dataset, achieving a commendable macro AUC score of 0.74. It is worth stressing that the training and validation happened on crop level but the testing on subject level. By averaging the prediction of all crops belonging to a patient, a subject-level prediction was formed.

\begin{table}[!htt]
\centering
\caption{
{\bf XResNet50 performance on different configurations.}}
\begin{tabular}{|r|l|l|}
\hline
 & Balanced & Unbalanced \\ \hline
\textbf{FL} & 0.71\space \scriptsize{(0.67, 0.75)} & \textbf{0.74}\space \scriptsize{(0.70, 0.77)}\\ \hline
CEL & 0.62\space\scriptsize{(0.57, 0.67)} & 0.65\space\scriptsize{(0.60, 0.70)}\\ \hline
\end{tabular}
\label{tab:xresnet-aucs}
\end{table}

\subsubsection{Comparative assessment}
When comparing the results from both models it is interesting to see that both reach a comparable performance despite fundamentally different input representations and model architectures. The most direct comparison is between the XGBoost model operating on SR (short-range) and the XResNet model, which by construction only leverages short-range features as well. It reveals a slight advantage on the side of the XResNet model, which is in line with the original hypothesis that the raw waveform contains additional discriminative information that is not covered in conventionally considered short-range ECG features. The long-range information contained in the HRV features is somewhat complementary to this information as the increase in the predictive performance of the HRV+SR model compared to the SR model shows. It remains an interesting question for future research if such long-range interactions could also be exploited using models operating on raw time series data with appropriate model architectures, see \cite{Mehari2023S4} for first steps in this direction.

\begin{figure}
    \centering
    \includegraphics[width=\columnwidth]{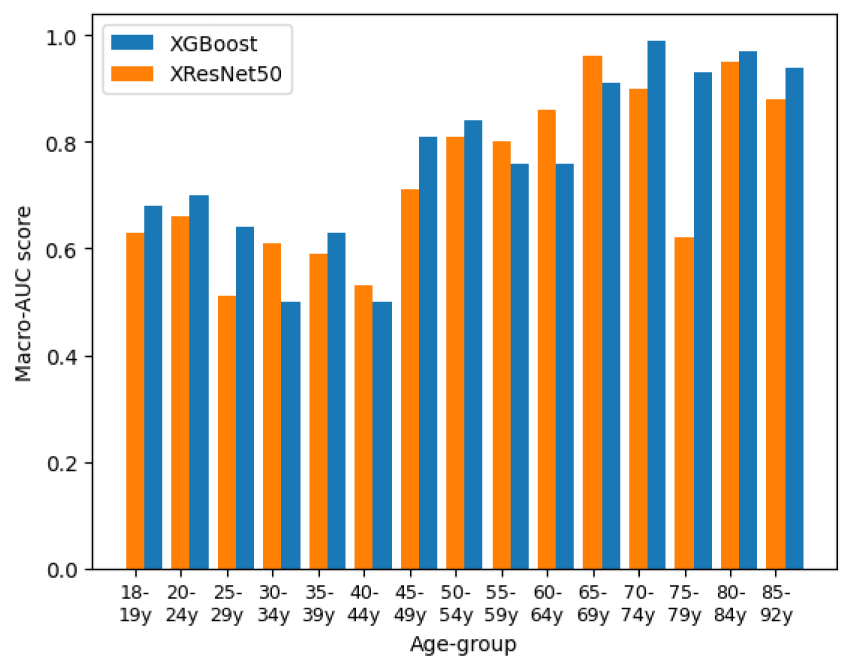}
    \caption{{\bf Predictive performance.}
Predictive performance results of the models per age group in terms of AUC on the test set, where the yellow (left) age group represents the XGBoost and the right (blue) the XResNet.}
    \label{fig:2}
\end{figure}

Additionally, Fig \ref{fig:2} displays sample-level AUC scores for both models across diverse age groups, revealing a general performance improvement with increasing age, albeit with a notable exception in the 75-79-year age group for the XResNet model and lower scores observed within the 24 to 44-year age range. These findings provide valuable insights into the nuanced age-related patterns discerned by both models.

\subsection{Explainability results}

\subsubsection{XGBoost}

\begin{figure*}
    \centering
    \includegraphics[width=\textwidth, height=0.8\textheight]{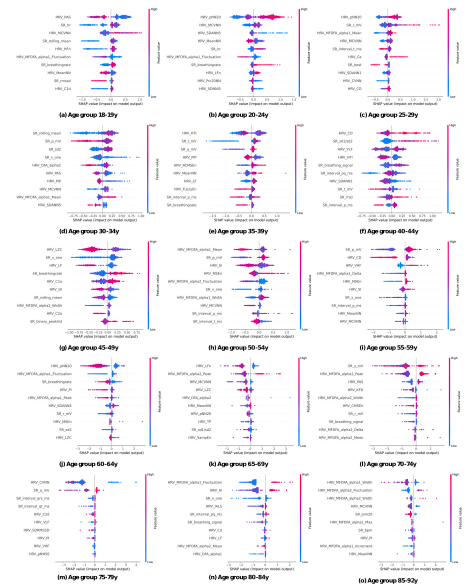}
    \caption{{\bf SHAP values across age groups.}
The 10 most important features for classifying the age groups are depicted in these subplots. In each subplot, the features are arranged in descending order of importance, emphasizing their significance in age group classification. The color scheme, with blue dots representing low feature values and red dots denoting high feature values, provides a visual representation of the feature’s influence across different age groups. As you move from left to right and from top to bottom, you explore the SHAP values for all age groups.}
\label{fig:3}
\end{figure*}

\heading{SHAP feature relevances} In this study, we classify the features into two categories: those with the prefix HRV- denote long-range HRV features, while those with the prefix SR- represent short-range features. Subsequently, we analyze the top 10 influential features for each age group, see Fig~\ref{fig:3}. 

\heading{Important features with consistent age trends} Observations reveal that certain features are recurrent across multiple age groups, displaying a consistent trend with respect to age. Specifically, HRV\_SDANN5, HRV\_PAS, P-wave amplitude (p\_mV) and alpha-fluctuation values consistently increase with age where lower values of these features are indicative of younger individuals, whereas higher values are associated with elderly individuals. Conversely, certain other features exhibit a contrasting trend. For instance, pNN20, MCVNN, and breathing rate in conjunction with breathing signal exhibit a decline with advancing age. High values of these features correspond to younger individuals, while lower values are characteristic of older individuals. Note that in the case of breathing rate and breathing signal, these features are derived from the ECG and serve as estimates of respiratory activity.

\heading{Consistency with literature results} It is noteworthy that our findings align with existing research in several aspects. Specifically, the observed trends in pNN, alpha-mean, HRV-PAS, P-wave amplitude and alpha-fluctuation are consistent with previous studies. For instance, the decrease in pNN50 with age among healthy subjects, as well as the discriminative power of pNN50 and pNN20 in age separation, has been noted by \cite{Mietus2002}. Similarly, the steady increase in alpha-values with age among healthy subjects, as well as rising alpha-fluctuations, has been reported \cite{BarqueroPerez2008} and \cite{Iyengar1996}, albeit without specific reference to alpha-mean. Furthermore, the upward trajectory of HRV-PAS with age as observed in our model, is consistent with the findings of \cite{Costa2017}. Similarly, the pattern of P-wave amplitude rising up to age 60 before declining, as observed in our study, concurs with the research of \cite{Havmoller2007}. In summary, our XGBoost model's conclusions are in accordance with existing research, reinforcing the notion that certain physiological features exhibit consistent age-related trends, which can be valuable in understanding the physiological changes associated with aging.

\heading{New insights: breathing rate and SDANN5} At this stage we have presented parts of our findings that align with previous research, however, we further provide insights into ECG and healthy aging, specifically for breathing rate and SDANN trends. 

According to \cite{Takayama2019} breathing rate and age are hardly correlated at all. However, the 2.5-97.5 percentile of the breathing rate increases with age according to \cite{RodrguezMolinero2013}, meaning that lower- and higher breathing rates become more common with age. Furthermore, \cite{schwartz1991aging} suggests that there are variations in respiratory dynamics, particularly in response to metronome breathing such as the increase in high frequency at different postural changes, especially in young subjects. In contrast to this work, these studies were not limited to healthy subjects only. Since all subjects in this work are healthy this means that the breathing rate decreases with age for healthy subjects. This finding is also plausible when considering that all subjects were in a resting state during recording. Because of the general decline in body activity with age less energy and thus oxygen is required to run body activities in a resting state. Assuming that the lungs and heart are healthy a lower breathing rate is therefore plausible for healthy aging. 

Our model reveals interesting insights regarding the SDANN5 feature. Contrary to established research showing a general decline in SDANN with age, the explainability analysis suggests that high SDANN5-values contribute positively towards the age prediction of older individuals. For instance, it associates lower SDANN5 values with those aged 20-34 and higher values with those aged 60-64. A similar study using the same dataset \cite{similar_paper} shows that while the mean SDANN does decline with age, there are significant variations in SDANN values among age groups, which lead to very high SDANN values being more probable for subjects older than 50 compared to subjects younger than 30. In summary, our XGBoost model uncovers an unexpected relationship between age and SDANN5, challenging the conventional wisdom of decreasing SDANN values with age. Notably, this effect is primarily observed in specific age groups beyond age 60.

\subsubsection{XResNet}

\begin{figure}
    \centering
    \includegraphics[width=\columnwidth]{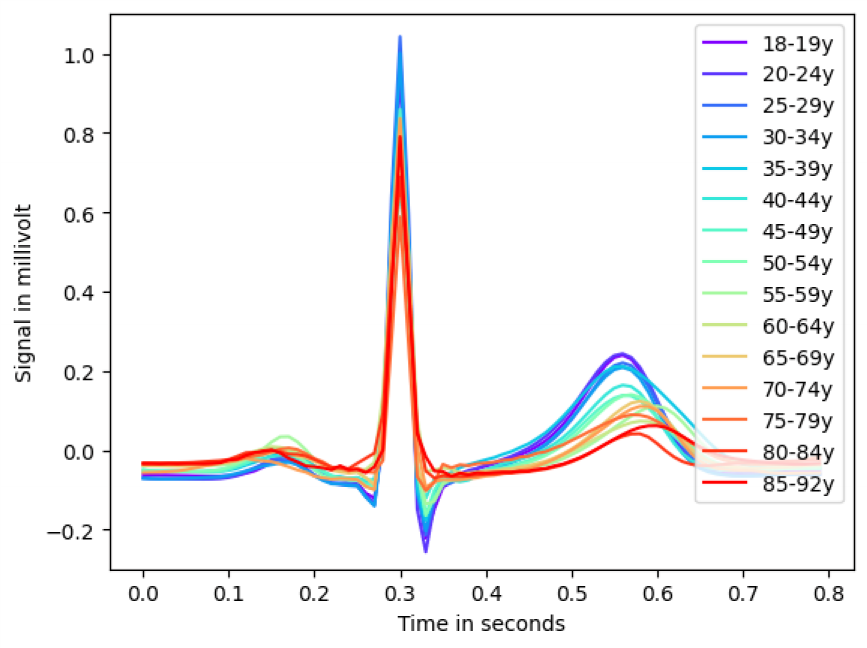}
    \caption{{\bf Aggregated mean heartbeat.}
Aggregated mean heartbeat for all age groups showcases ECG feature trends across age groups.}
\label{fig:4}
\end{figure}

\heading{Beat-level descriptive analysis} At first, we explore superimposed mean heartbeats for all age groups in Fig~\ref{fig:4} as a plausibility test and to compare with literature statements. The amplitude of the T-wave decreases with age and shifts to the right, indicating an overall longer cardiac cycle, meaning a slower heart rate. Furthermore, the T- and P-wave intervals shorten with age; moreover, the absolute magnitude of the S-peak, Q-peak, and P-wave appears to diminish with age as well, which is in accordance with \cite{Zareba2020}\cite{Bocchi2020}. It is noteworthy that the amplitude of the R-peak shows no conclusive trend with age.

\begin{figure*}
    \centering
    \includegraphics[width=\textwidth, height=0.8\textheight]{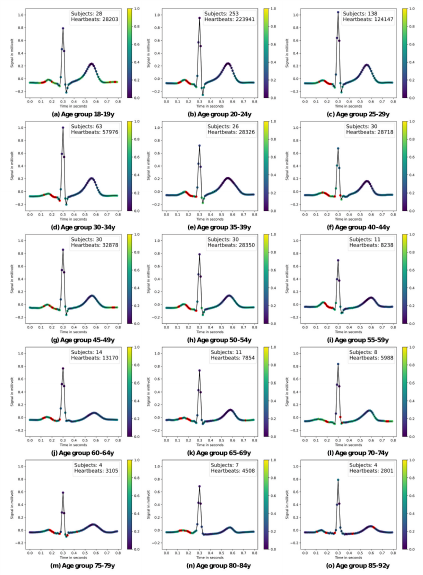}
    \caption{{\bf Saliency maps across age groups.}
Within each subplot, as you progress from left to right and from top to bottom, you navigate through the beat-level ECG saliency maps, shedding light on the key ECG features that contribute to age group differentiation according to the color map scheme. ’Subjects’ and ’Heartbeats’ state the number of subjects and heartbeats used to create these plots. The eight highest gradients are marked in red.}
    \label{fig:5}
\end{figure*}

\heading{Aggregated saliency maps: methodology}
Since ECGs even in the same age group have slightly different heart rates and are generally not aligned, the crop-level-saliency maps cannot simply be laid on top of each other. Following \cite{wagner2023explaining}, the crops of each subject with their saliency maps were split into individual heartbeats and averaged from 30 milliseconds before to 50 milliseconds after the R-peak. Then, these medium heartbeats were again averaged for each age group, resulting in one aggregated heartbeat per age group as shown in Fig~\ref{fig:5}. To reveal the patterns exploited by the model most clearly, we used the training set to produce the aggregated attribution maps. We also mark the most salient data points (marked in red) to identify patterns across age groups as described below.

\heading{Aggregated saliency maps: results} The XResNet model consistently demonstrates a predilection for the entire P-wave as individuals age. It specifically focuses on the offset, with some onset in early age groups. These variations may reflect different P-wave types. Prior studies found that, the distribution of which undergoes significant changes with age. \cite{Havmoller2007}. Furthermore, research by \cite{jcm11133737,Turhan2003,lindow2022heart} has elucidated age-related disparities in various aspects of the P-wave, including its duration. Consequently, it is plausible that the Deep Learning model distinguishes age groups based on distinct P-wave parts and their respective distributions, underlining the complexity of its age classification methodology. The application of more sophisticated methods, for example from the domain of concept-based XAI such as \cite{Vielhaben:2022MCD}, would be the logical next step to uncover these changes. Apart from the P-wave, the model frequently focuses on the Q-peak and S-peak while showing limited relevance to the R-peak and the peak of the T-wave. The TP segment receives moderate attention, indicating its importance in age-related classification.

\subsubsection{Comparative assessment}

Table \ref{tab:explainability-results} presents key insights into age-related trends of ECG features that were derived from applying XAI on the XGBoost and XResNet50 models. In the XGBoost model, age is associated with a decrease in pNN20, MCVNN, and breathing rate, along with an increase in alpha-fluctuations, P-wave amplitude, and PAS. Additionally, SDANN5 values rise with age. In contrast, the XResNet50 exhibits distinct focus areas with age: given our criterion over the eight most important saliency time steps, it emphasizes P-wave features (18.33\% for onsets and 53.33\% for offsets), frequently places relevance on the Q-peak (8.33\%), shows little relevance on the  R-peak, sometimes focuses on the S-peak (4.16\%), attributes minimal relevance to the T-wave (3.33\%), and frequently assesses the TP-interval (12.5\%). The latter might be related to differences in the heart rate, which are difficult to analyze by means of saliency maps. Nevertheless, the observed trends offer valuable insights into age group differentiation in the two models' ECG interpretations.

\begin{table}[!htt]
\centering
\caption{
{\bf Clinical observations and feature trends in age-related explainability. Percentages refer to relative number of age groups where a high-saliency timestep (red in Fig~\ref{fig:5}) occurred in the corresponding segment.}}
\begin{tabular}{|p{0.35\linewidth}|p{0.55\linewidth}|}
\hline
XGBoost feature trends & XResNet50 observations \\ \hline
pNN20 $\downarrow$ & Strong focus on P-wave offsets (53.33\%), with some P-onsets (18.33\%) in early age groups. \\ \hline
Alpha-fluctuations $\uparrow$ & Focus on Q-peaks (8.33\%), especially in middle-aged groups.  \\ \hline
MCVNN $\downarrow$   &  R-peak is largely ignored by the model. \\ \hline
P-wave amplitude $\uparrow$  & S-peak (4.16\%) is important for age groups 35-44y and 70-79y. \\ \hline
PAS $\uparrow$  & Hardly any focus on T-wave (3.33\%). \\ \hline
breathing rate and signal $\downarrow$ & Frequent focus on TP-segment (12.5\%). \\ \hline
SDANN5 $\uparrow$ &   \\ \hline
\end{tabular}
\label{tab:explainability-results}
\end{table}

\subsubsection{Data imbalance and research focus} While the 'autonomic aging' dataset \cite{dataset_source} used in this work stands out as one of the largest datasets of its kind, it is important to acknowledge its inherent imbalance, notably the scarcity of samples from individuals aged 70 or older. Deep learning models, with their appetite for ample training data, face a particular challenge in such scenarios. Addressing the imbalance by consolidating the underrepresented age groups might seem like a logical step, however, leads to a less nuanced prediction model. We have deliberately chosen not to merge age groups above 70 years as our primary focus centers on understanding the nuances of a healthily aging heart. In this context, we find that the uniqueness of our dataset, even with its imbalances, continues to yield more insightful results that better align with our research objectives.

\section{Conclusion}

In this study, we investigated age-related cardiovascular changes of a healthy population. Leveraging ECG data from 1,095 healthy subjects, we developed two different models that work across diverse data modalities and used feature attribution methods to study their behavior: an XGBoost model analyzing short- and long-range ECG features as well as an XResNet50 model processing raw ECG data. Our experiments suggest that the feature-based model achieves better predictive performance in comparison with raw ECG data, which is comparable with literature performance. The findings from the feature-based model indicate increasing heart irregularity and reduced flexibility with age, which aligns with prior research. It also revealed a decline in inferred breathing rate with age and the significance of high SDANN values in older individuals. Notably, the deep-learning model identified the P-wave as the most important segment across all age groups. Our findings provide complementary insights into age-related ECG changes, whose identification is crucial for the early detection of cardiovascular diseases. To promote further exploration in this area of study, we release the source code underlying our study \cite{Ott_ECG-aging_2023}.

\end{document}